\documentclass[aps,prl,twocolumn,showpacs,superscriptaddress,floatfix]{revtex4-2}
\usepackage[utf8]{inputenc} 
\usepackage{amssymb,amsmath}
\usepackage{graphicx}
\usepackage{braket}
\usepackage{siunitx}
\usepackage{xcolor}
\usepackage{gensymb} 
\usepackage{bbold}
\usepackage{physics}
\usepackage{verbatim}

\def\lambdar{\lambda_{\rm R}}                            		
\def\kr{k_{\rm R}}                            				
\def\Er{E_{\rm R}}                            				
\def\Rb87{^{87}\mathrm{Rb}}                             
\def\K40{^{40}\mathrm{K}}                    		    

\def\us{\mu \text{s}}

\def\ex{\mathbf{e}_x}  
  
\def\ey{\mathbf{e}_y}  
\def\ez{\mathbf{e}_z}  

\def\phicolorI{teal}
\def\phicolorII{magenta}

\newcommand{\Fx}{\langle \hat{F}_x \rangle}

\newcommand{\expx}{\expectationvalue{\hat\sigma_x}}
\newcommand{\expy}{\expectationvalue{\hat\sigma_y}}
\newcommand{\expz}{\expectationvalue{\hat\sigma_z}}

\newcommand{\expxq}{\expectationvalue{\hat\sigma_x(q)}}
\newcommand{\expyq}{\expectationvalue{\hat\sigma_y(q)}}
\newcommand{\expzq}{\expectationvalue{\hat\sigma_z(q)}}

\newcommand{\sigmaq}{\langle \hat {\boldsymbol \sigma}(q)\rangle}

\newcommand{\rfph}{\phi_{\text{rf}}}
\newcommand{\Omrf}{\Omega_{\text{rf}}}

\usepackage[normalem]{ulem}

\begin{document}

\title{Observation of dynamical topology in 1D}
\author{G.~H.~Reid}
\author{Mingwu Lu}
\author{A.~R.~Fritsch}
\author{A.~M.~Pi\~{n}eiro}
\author{I.~B.~Spielman}
\affiliation{Joint Quantum Institute, National Institute of Standards and Technology, and University of Maryland, Gaithersburg, Maryland, 20899, USA}
\email{spielman@nist.gov}
\date{\today}
\begin{abstract}
Nontrivial topology in lattices is characterized by invariants---such as the Zak phase for one-dimensional (1D) lattices---derived from wave functions covering the Brillouin zone.
We realized the 1D bipartite Rice--Mele (RM) lattice using ultracold $\Rb87$ and focus on lattice configurations possessing various combinations of chiral, time-reversal and particle-hole symmetries.
We quenched between configurations and used a form of quantum state tomography, enabled by diabatically tuning lattice parameters, to directly follow the time evolution of the Zak phase as well as a chiral winding number.
The Zak phase evolves continuously; however, when chiral symmetry transiently appears in the out-of-equilibrium system, the chiral winding number is well defined and can take on different integer values.
When quenching between two configurations obeying all three symmetries the Zak phase is time independent; we confirm the contrasting prediction of [M. McGinley and N. R.Cooper, PRL {\bf 121} 090401 (2018)] that chiral symmetry is periodically restored, at which times the winding number changes by $\pm 2$, yielding values that are not present in the native RM Hamiltonian.
\end{abstract}
\maketitle

Topological invariants robustly classify gapped quantum systems in equilibrium.
In addition to dimensionality, the presence or absence of symmetries determines the topological invariants that characterize them~\cite{schnyder2008classification, kitaev2009periodic, chiu2016classification}.
Thus, these invariants remain constant provided that no gaps close and reopen and no symmetries are added or removed.
For this reason, one might expect the topology of dynamical quantum systems to be similarly robust; this expectation is untrue.
We experimentally study the dynamical topology of ultracold atoms in a one-dimensional (1D) bipartite lattices in terms of the Zak phase~\cite{zak1989berry} and chiral winding number.
As predicted by Ref.~\cite{mcginley2018topology}, we observe that these quantities evolve in time depending on how symmetries change between the initial state and the evolution Hamiltonian.

Despite their relative simplicity, 1D bipartite lattices have nontrivial topology~\cite{asboth2016topo} characterized by the Zak phase
\begin{equation}
    \phi_{\rm Z} = i \int_{\rm BZ} d q \bra{\psi(q)}\partial_q\ket{\psi(q)},\label{eq:zak}
\end{equation}
a Berry's phase~\cite{berry1984phase} resulting from the distribution of crystal momentum states $\ket{\psi(q)}$ throughout the Brillouin zone (BZ).
We implemented a general approach for obtaining the Zak phase by measuring the wave function through a form of quantum state tomography~\cite{alba2011ToF} and directly evaluating Eq.~\eqref{eq:zak}. 
The Zak phase has been previously measured using a special purpose interferometric technique with cold atom~\cite{atala2013direct} and in photonic systems~\cite{wang2016measurement, cardano2017detection} and resonant circuits~\cite{goren2018topological}.

\begin{figure}[tb!]
\includegraphics{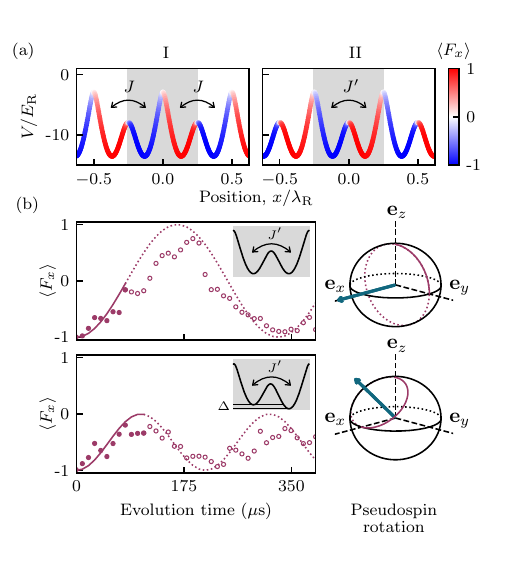}
\caption{(a) Adiabatic potentials colored according to the local magnetization for configurations I and II with the $j=0$ unit cell marked in gray, and computed for $[\Omega_+ = 13.2 \Er, \Omega_- = 5.5 \Er, \Omrf = 3.5 \Er]$.
(b) Pseudospin evolution in the ``$y$'' and ``$x$'' readout lattices.
The left panels plot $\Fx(t)$ evolving according to the lattice pictured in its inset (experiment: symbols, theory: solid curve), times up to the rotation time for the readout lattice are shown with solid symbols/curves, while evolution after is shown with open symbols/dotted curves. 
The right panels show the computed evolution on the Bloch sphere for these trajectories along with the axis of rotation in blue.
}\label{fig_setup}
\end{figure}

We employed a bipartite optical lattice~\cite{lu2016geometrical,Lu2022Dirac} to realize the Rice--Mele (RM) Hamiltonian~\cite{rice1982elementary}
\begin{align}
    \hat{H}_{\text{RM}} = \sum_j & \Big[-\big( J' \dyad{j}{j}+ J\dyad{j+1}{j}\big)\otimes\dyad{\downarrow}{\uparrow} +\text{h.c.} \nonumber \\
     &+ \Delta \dyad{j}{j} \otimes \hat \sigma_z \Big],
    \label{eq:RM}
\end{align}
where $j$ labels the unit cell; $\uparrow$ and $\downarrow$ identify sub-lattice sites that we associated with a pseudospin degree of freedom; and $J$ and $J'$ are the inter- and intra-cell tunneling strengths respectively.
For the special case of $\Delta = 0$ the RM Hamiltonian reduces to the highly symmetric Su--Schrieffer--Heeger (SSH) Hamiltonian~\cite{su1979solitons}.
As shown in Fig.~\ref{fig_setup}(a), we label lattices with $\Delta = 0$ and $J \gg J'$ as configuration I (topologically non-trivial with $\phi_{\rm Z}=\pi$), and those with $\Delta = 0$ and $J' \gg J$ as configuration II (topologically trivial with $\phi_{\rm Z}=0$).
The SSH Hamiltonian obeys a chiral sublattice symmetry (CS) $\hat S\hat H \hat S^\dagger = -\hat H$ with $\hat S = \hat\sigma_z$, a particle-hole symmetry (PHS) $\hat C \hat H^* \hat C^\dagger = -\hat H$ with $\hat C = \hat\sigma_z$, as well as time-reversal symmetry (TRS)  ~\footnote{Our expressions for the symmetry operations are the unitary part of the complete symmetry operator, where complex conjugation implicitly expresses the antiunitary contribution}.

In the RM model (with only TRS), the Zak phase is real valued and defined ${\rm mod}\ 2\pi$; for the SSH model (with all three symmetries) the constraint on the Zak phase changes to $\phi_{\rm Z}/\pi = \nu$, defining an integer valued winding number $\nu$.
Here, CS constrains the eigenstates of Eq.~\eqref{eq:RM} to reside in the equatorial plane of the Bloch sphere, and the topological invariant $\nu$ counts the number of times the state encircles the equator as $q$ traverses the BZ~\cite{asboth2016topo, cooper2019topological, NuFootnote}.

We experimentally and numerically study initial states characterized by two different symmetries and evolve them according to the SSH Hamiltonian in both the trivial and topological configurations.
When PHS and CS are absent in the initial state, $\phi_{\rm Z}$ evolves in time, except when $J = 0$~\footnote{As discussed in the SM, when TRS and PHS are absent in the initial state (but CS is present) $\phi_{\rm Z}$ also can evolve in time.}.
When no symmetries change $\phi_{\rm Z}$ is time independent; however, in agreement with Ref.~\onlinecite{mcginley2018topology}, the time-evolving state periodically recovers CS at times where the winding number $\nu$ becomes well defined.
Remarkably we observe $\nu$ alternating either between $-1$ and $+1$ or between $0$ and $-2$ even though the SSH model only allows winding numbers $0$ and $\pm 1$.

{\it Experimental System} Our experiments began with harmonically trapped $\Rb87$ Bose-Einstein condensates [BECs, with trap frequencies $(f_x,f_y,f_z) \approx (25, 150, 100)\ {\rm Hz}$] in the $\ket{f=1, m_F=-1}$ hyperfine state, that experienced two ``Raman'' fields (generated by lasers with wavelength $\lambdar\approx790\ {\rm nm}$ and Rabi frequencies $\Omega_+$ and $\Omega_-$) and one radiofrequency (rf) magnetic field (with Rabi frequency $\Omrf$ and phase $\rfph$ with respect to the Raman fields), coupling the $\ket{f=1, m_f = 0,\pm 1}$ hyperfine states.
The single photon recoil wave-vector $\kr = 2 \pi /\lambdar$ and energy $\Er = \hbar^2 \kr^2/ 2m$ specify the natural momentum and energy scales of this system.
This combination of Raman and rf fields approximates the RM Hamiltonian derived from the adiabatic potentials in Fig. \ref{fig_setup}(a), with unit cell sized $\lambdar/2$.
In practice $\rfph$ controlled both the energy splitting $\Delta$ as well as selected the SSH configuration~\footnote{Our calibrated values for $\rfph$ differed by $~3\%$ from the model prediction indicating small deviations from our model.}, and $\Omrf$ controlled relative strength of $J$ and $J^\prime$.

We loaded the BEC into the lattice in a three step process.
First, the rf and Raman fields ramped on in 2.5 ms, transferring the BEC to the lowest band of an initialization lattice.
Second, we dephased the BEC to fill the BZ~\cite{pineiro2019sauter} and lastly we switched to the final SSH configuration.

We measured the pseudospin resolved momentum distribution by combining time-of-flight (ToF) imaging with a form of quantum state tomography.
The $\uparrow$ and $\downarrow$ sites are highly polarized, with $\Fx\approx\pm1$, corresponding to atomic states $\ket{m_x = \pm1}$ as indicted by the coloration in Fig. \ref{fig_setup}(a). 
Our ``default'' readout sequence began by removing the coupling fields and applying an rf pulse to map eigenstates of $\hat  F_x$ to our $\hat F_z$ measurement basis. 
During the following ToF a magnetic field gradient separated the hyperfine states by the Stern-Gerlach effect, ultimately yielding the momentum distribution of the $\ket{\uparrow,\downarrow}$ pseudospin states.
Summing the populations separated by $2\kr$ yielded the pseudospin resolved crystal momentum distribution, from which we obtained $\expzq$.

We found $\expxq$ and $\expyq$ by evolving our system under one of two readout lattices depicted in Fig.~\ref{fig_setup}(b) prior to this sequence.
The ``$y$'' readout lattice shown in (b) is described by a RM model with $J=\Delta=0$; unitary evolution under this lattice implements pseudo-spin rotations about $\hat \sigma_x$.
Evolving for a $\pi/2$ time (vertical line) transforms $\ket{y_\pm}$ to $\ket{\uparrow\downarrow}$ (right).
The ``$x$'' readout lattice in (b) uses $\Delta\approx J^\prime$, and similarly transformed $\ket{x\pm}$ to $\ket{\uparrow\downarrow}$. 
Combining these three complimentary measurements allowed us to reconstruct the pseudospin Bloch vector $\langle\hat{\boldsymbol \sigma}(q)\rangle\equiv\left\{\expxq, \expyq, \expzq\right\}$.

Quantum state tomography reconstructs the density operator from a set of measured expectations values, here ${\boldsymbol \sigma}(q)$ suffices. 
In general our measurement has imperfect contrast, reducing the purity of the reconstructed density operator. To compare with the predicted pure states, we compute the pure state that most closely matches our experimental result by taking the principal eigenvector of the density operator (equivalent to normalizing $\sigmaq$), in the spirit of Ref. \onlinecite{diehl2011topology}.
\footnote{This process leaves the overall phase undefined. The choice of phase can be viewed as a momentum space gauge transformation which affects no observables, but may introduce an overall offset to the calculated Zak phase.}

We validated our pseudospin measurement method by obtaining $\sigmaq$ for ground band eigenstates in both configurations.
The momentum-space Hamiltonian $ \hat H({q}) = -{\boldsymbol h}( q)\cdot \hat {\boldsymbol \sigma}$ is expressed in terms of a polarizing field
\begin{equation}
{\boldsymbol h}(q) = \left[J^\prime + J\cos\left(\frac{\pi q}{\kr}\right), J \sin\left(\frac{\pi q}{\kr}\right), \Delta\right] \label{eq:ssh_st}
\end{equation}
defining the axis along which the eigenstates are aligned.
In limiting cases of configurations I and II for the SSH model, these axes are $\left[\cos (\pi q/\kr), \sin (\pi q/\kr), 0\right]$ and $\left[1,0,0\right]$ respectively.

\begin{figure}[t]
\includegraphics[width=\columnwidth]{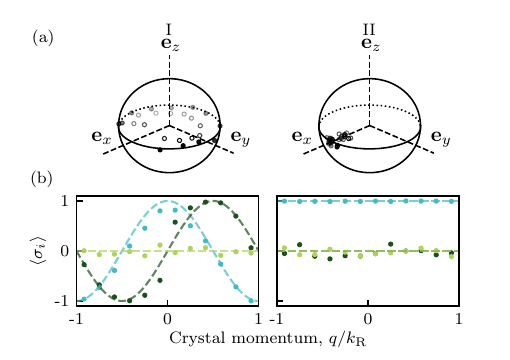}
\caption{
Pseudospin makeup of  ground band in SSH configurations I and II.
(a) pseudospin state for $q$ values sampling the whole BZ plotted in the Bloch sphere. 
The raw measurements (open symbols) are impure (i.e., magnitude $<1$); the solid symbols mark the nearest pure states on the surface of the Bloch sphere.
For the eigenstates of configuration I the vectors trace the equator of the Bloch sphere, while for configuration II they are aligned along $\ex$. 
(b) pure-state expectations values of $\hat\sigma_x, \hat\sigma_y, \hat\sigma_z$ shown as a function of $q$ in light blue, dark green, and light green, plotted along with theory (dashed curves). 
\label{figSSH}} 
\end{figure}

Figure~\ref{figSSH}(a) compares our observation of $\langle\hat{\boldsymbol \sigma}(q) \rangle$ (bold symbols) with the prediction of the SSH model for configuration I; the top panel renders these data as points on the Bloch sphere and the bottom panel explicitly plots the components of $\langle\hat{\boldsymbol \sigma}(q)\rangle$. 
For this configuration, the eigenstates encircle the equator of the Bloch sphere as $q$ ranges $-\kr$ to $+\kr$, and the SSH model predicts a Zak phase $\phi_{\rm Z} = \pi$.
By contrast panel (b) shows configuration II; the eigenstates are $q$ independent, which predicts $\phi_{\rm Z} = 0$.
In both cases, hollow symbols inside the Bloch sphere display our measurements prior to taking the principal eigenvector of the density operator, showing reduction purity from the combination of imperfect measurement and state preparation.

We experimentally obtained $\phi_{\rm Z}$ from discretely sampled $q$ values inspired by the technique introduced in Ref.~\onlinecite{fukui2005chern}.
In good agreement with the theory this gives $0.99 (3) \pi$ and $-0.0005 (1) \pi$ for configurations I and II respectively~\footnote{The stated uncertainties are the sample standard deviation of Zak phase calculated individually from about 30 separate measurements.}; for I, the deviation from unity was likely caused by a combination of residual $\Delta$ and systematic counting bias in our image analysis.

{\it Topology out of equilibrium}
Having measured the Zak phase of eigenstates of the SSH model, we turn to the dynamics of initial states characterized by different symmetries and evolve them under the SSH Hamiltonian which respects all three symmetries (CS, PHS and TRS).

We first focus on the $q$-independent initial state $\ket{\psi(q)} = \ket{\downarrow}$, the ground state of the $\Delta\gg (J,J^\prime)$ RM Hamiltonian that only retains TRS.
This initial state was prepared by adiabatically loading into a maximally imbalanced initialization lattice with $\Delta \approx 5 \Er$ and $J=J^\prime\approx 0.1\ \Er$.
We initialized evolution by abruptly switching to a maximally dimerized lattice with $\Delta=0$ in either configuration I or II.

\begin{figure}[t]
\includegraphics{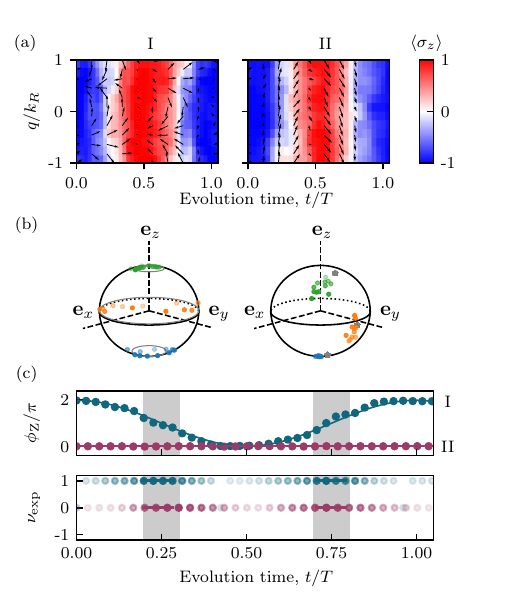}
\caption{Momentum resolved pseudospin evolution in SSH configurations I and II. Model parameters were $J = 0.395 (2) \Er$, $J^\prime = 0.012 (2) \Er$  for configuration I and $J = 0.038 (3) \Er$, $J^\prime = 0.379 (2) \Er$  for configuration II.
(a) Reconstructed Bloch vectors. 
Color represents $\expzq$ while black arrows denote $(\expxq, \expyq)$.
The data were filtered in crystal momentum and time to improve the signal-to-noise ratio (with root mean square Gaussian widths $\kr/8$ and $10 \us$). 
(b) Corresponding points on Bloch sphere for evolution times of $(0.05, 0.25,0.45)\ T$. 
(c) Zak phase and winding number for configurations I (\phicolorI) and II (\phicolorII). 
The color's intensity indicates the extent to which the measured state breaks CS and the grey boxes surround the times when CS is predicted to be recovered, with width $T/10$.
\label{tunnel_fig}}
\end{figure}

In both configurations chiral symmetry implies that ${\bf h}(q)$ is in the $\ex$-$\ey$ plane, and owing to the nearly flat bands of the highly dimerized SSH Hamiltonian $|{\bf h}(q)|$ is almost constant.
As a result, $\langle\hat\sigma_z\rangle$ exhibits nearly $q$-independent full contrast oscillations for the $\ket{\downarrow}$ initial state as shown in Fig.~\ref{tunnel_fig}(a).
The $\langle\hat\sigma_{x,y}\rangle$ components (black arrows) evolve in a $q$-dependent way in configuration I but are $q$-independent in configuration II.

In configuration I each $q$ state orbits about a different axis in $\ex$-$\ey$ plane, causing $\sigmaq$ to first spread, then encircle the Bloch sphere, before ascending to converge at $\ket{\uparrow}$.
The corresponding evolution of the Zak phase [\phicolorI{} points in Fig.~\ref{tunnel_fig}(c)] starts at $2 \pi$ (equivalent to 0, modulo $2 \pi$) and reaches $\pi$ when  $\sigmaq$ reaches at the equator.
When $\sigmaq=\ez$ the Zak phase reaches its extremal value of $0$. The state continues to evolve, returning to the initial configuration at $T \approx 360\ \us$.

In configuration II ${\bf h}(q) = [J^\prime,0,0]$, as a result $\sigmaq$ orbits around $\ex$, independent of $q$, starting from the $-\ez$ pole and reaching the $+\ez$ pole via $\ey$ and returning to $-\ez$.
The derivative in Eq.~\eqref{eq:zak} implies $\phi_{\rm Z}=0$ at all times, in agreement with our observations [\phicolorII{} points in Fig.~\ref{tunnel_fig}(c)].
More generally, the SSH Hamiltonian (with $J,J^\prime\neq0$) has $q$-dependent evolution making the Zak phase time-dependent except when $J=0$.
In both configurations the time-evolving state is always an eigenstate of some RM Hamiltonian in the initial configuration, but with time dependent $\Delta$ and complex tunneling.
For most of the evolution, the state is described by a RM model violating all symmetries (non-topological symmetry class A).
Twice every oscillation $\sigmaq$ aligns along $\ez$, at which times the state obeys TRS but violates CS (non-topological symmetry class AI).
Similarly the $\sigmaq$ lies on the equator twice per oscillation and the system becomes an eigenstate of the SSH model with complex tunneling phase $\phi=\pm\pi/2$, thereby recovering CS but violating TRS (topological symmetry class AIII).
The recovered chiral symmetry makes the winding number well defined, with $\nu=1$ in configuration I and $\nu=0$ in configuration II.

Because all experimental data has some contribution along $\ez$ and therefore violates CS to some degree, we projected the measurements onto the $\ex$-$\ey$ plane, enforcing CS, before computing an integer valued winding number $\nu_{\rm exp}$ [bottom panel of Fig.~\ref{tunnel_fig}(c)].
The intensity of each symbol marks the projection of the reconstructed state onto the $\ex$-$\ey$ plane, whereby bold symbols mark states with little violation of CS. 
The vertical grey bands mark the regions within $10
\%$ of $T/4+nT/2$, the times when CS is expected to be recovered.
$\nu_{\rm exp}$ is defined at all times but should only be compared to $\nu$ when CS is recovered. 
We see that within the grey bands the CS is maximally restored (bold symbols) and we confirm $\nu_{\rm exp} = \nu$ for these times.
Interestingly, $\nu_{\rm exp}$ continues to agree with the predicted value of $\nu$ even when CS is violated, except for three points in configuration I when $\langle \hat {\boldsymbol \sigma}(q)\rangle$ is everywhere nearly aligned along $\ez$ and consequently measurement noise dominates the $\ex$-$\ey$ projection.

{\it Dynamically induced symmetry breaking}
We conclude by investigating cases where the initial state and evolution Hamiltonian respect all three symmetries---TRS, PHS and CS---by preparing eigenstates of the fully dimerized SSH Hamiltonian and then evolving under the opposite SSH configuration.
As one might expect, the Zak phase is predicted to be constant at all times; however, CS is lost during much of the evolution and when it is recovered $\nu$ can take on values that are not present in the SSH Hamiltonian, confirming a counterintuitive prediction of Ref.~\cite{mcginley2018topology}.

\begin{figure}[t]
\includegraphics{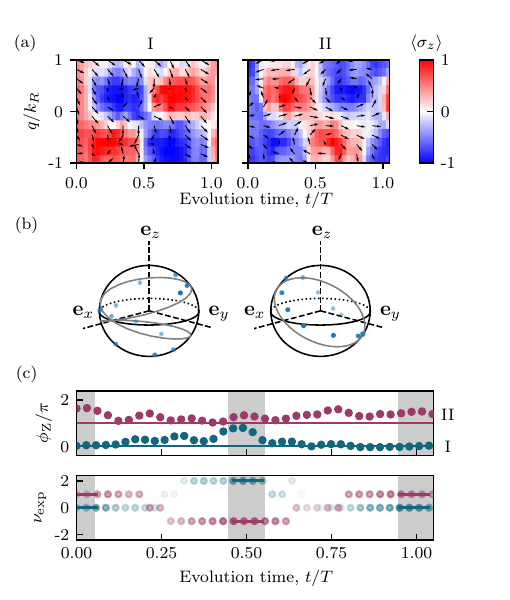}
\caption{Momentum resolved pseudospin evolution in SSH configurations I and II, using initial states of the opposite configuration. Model parameters were $J =0.408 (3) \Er$, $J^\prime =0.003 (5) \Er$  for configuration I and $J = 0.009 (5)  \Er$, $J^\prime =0.446 (3)\Er$  for configuration II.
All panels are plotted as in Fig.~\ref{tunnel_fig}.
(a) Individual expectation values $\expx, \expy, \expz$. (As before the data were filtered with root mean square Gaussian widths $\kr/8$ and $10 \us$.) 
(b) Bloch state rendering at times $t = 0.40  T$.
(c) Zak phase and winding number for configurations I (\phicolorI{}) and II (\phicolorII{}).
\label{fig_flip}} 
\end{figure}

As before, when the system evolves in configuration I, the state for each $q$ value orbits a different axis in $\ex$-$\ey$ plane giving the distributions in Fig. \ref{fig_flip}(a)-left.
In this case the initial state, a configuration II eigenstate, has $\langle \hat {\boldsymbol \sigma}(q)\rangle$ aligned along $\ex$ for all $q$. 
Figure~\ref{fig_flip}(b)-left shows that the time evolution of $\langle \hat {\boldsymbol \sigma}(q)\rangle$ traces out a figure-8 shape consisting of symmetric loops in the upper and the lower hemisphere of the Bloch sphere.
Ideally, the Zak phase is time independent since the two loops enclose equal areas but are traced in opposite directions as a function of $q$, thereby giving equal but opposite contributions to Eq.~\eqref{eq:zak}.
The data in Fig.~\ref{fig_flip}(c) is in qualitative agreement with this prediction.
Despite this, the topology of the state changes when the state recovers CS every $T/2=\pi/(2 J) \approx 160\ \us$ when $\nu$ alternates between $0$ and $2$ [Fig.~\ref{fig_flip}(c)], which is not possible for an SSH model eigenstate. 

Figure \ref{fig_flip}(a)-right shows evolution under configuration II, again resulting from a $q$-independent rotation around $\ex$.
In this case the initial state is a configuration I eigenstate where $\langle \hat {\boldsymbol \sigma}(q)\rangle$ fully encircles the equator of the Bloch sphere.
As time evolves, $\langle \hat {\boldsymbol \sigma}(q)\rangle$ describes a great circle rotated by an angle $2 J^\prime t$ about $\ex$ as shown in Fig.~\ref{fig_flip}(b)-right.
Since the circle always encloses half the area of the Bloch sphere, we expect $\phi_{\rm Z} = \pi$ for all time.
As in the previous case, the state periodically recovers CS when it returns to the equator, at which times the winding number alternates between $\nu = +1$ and $-1$.

The top panel of Fig.~\ref{fig_flip}(c) plots the time evolving Zak phase in configurations I and II in \phicolorI{} and \phicolorII{} respectively.
Both of these fluctuate near the expected value; we attribute these fluctuations to imperfections in state preparation, the evolution Hamiltonian and the infidelity in our readout process. 
As in Fig.~\ref{tunnel_fig} the grey bands mark the expected times when CS is restored and in agreement with our model, $\nu_{\rm exp}$ oscillates between 0 and 2 for configuration II and between +1 and -1 for configuration II.
In both of these cases, the Zak phase is ideally constant, while the winding number discontinuously changes.
This is possible because the Zak phase is defined modulo $2 \pi$ allowing for $\nu$ to change by multiples of 2 at constant $\phi_{\rm Z}$.

We note that while $\phi_{\rm Z}$ can be heavily affected by noise and imperfections as seen in Fig.~\ref{fig_flip}(c), $\nu_{\rm exp}$ is more robust, deviating from the prediction only when the noise is comparable in strength to the projected measurements.

{\it Discussion and outlook} We presented paradigmatic examples of how the topology of a quantum system can change during out-of-equilibrium evolution.
From a macroscopic perspective, the out-of-equilibrium evolution of the Zak phase is associated with a current between unit cells and the resulting change in polarization~\cite{mcginley2018topology}. 
In a fully dimerized configuration II case there is no current between unit cells and the Zak phase must be constant, while in configuration I the probability amplitude oscillates between $\ket{\uparrow}$ and $\ket{\downarrow}$ sites in adjacent unit cells and the Zak phase can change.
The associated physical displacement was directly observed in Ref.~\onlinecite{Lu2022Dirac} while observing Floquet topological invariants~\cite{Kitagawa2010}.

At these times when $\nu=2$, the system approaches an eigenstate of an extended SSH model where next-nearest neighbor tunneling dominates~\cite{hsu2020topological}.
This marks the ability of unitary evolution under relatively simple Hamiltonians to dynamically prepare eigenstates of experimentally inaccessible models.
A natural extension of this work is dynamical symmetry breaking and recovery for strongly correlated systems: when do similar concepts apply to interacting systems and what otherwise inaccessible eigenstates can be realized?

Li. et al. and Fläschner et. al. used different but conceptually related approaches to measure Berry phases and curvatures in 2D honeycomb lattices~\cite{li2016bloch, flaschner2016Berry}. based the form of Bloch state tomography proposed by Hauke et. al.~\cite{hauke2014tomography}. This was extended to out-of-equilibrium  states in Ref.~\onlinecite{flaschner2018observation}.

\begin{acknowledgments}
We thank M.~McGinley and N.~R.~Cooper for helpful discussions as well as N.~Pomata and M.~Doris for carefully reading our manuscript.
This work was partially supported by the National Institute of Standards and Technology, and the National Science Foundation through the Physics Frontier Center at the Joint Quantum Institute and the Quantum Leap Challenge Institute for Robust Quantum Simulation.
\end{acknowledgments} 

\bibliography{spin_tomo}

\begin{thebibliography}{32}%
\makeatletter
\providecommand \@ifxundefined [1]{%
 \@ifx{#1\undefined}
}%
\providecommand \@ifnum [1]{%
 \ifnum #1\expandafter \@firstoftwo
 \else \expandafter \@secondoftwo
 \fi
}%
\providecommand \@ifx [1]{%
 \ifx #1\expandafter \@firstoftwo
 \else \expandafter \@secondoftwo
 \fi
}%
\providecommand \natexlab [1]{#1}%
\providecommand \enquote  [1]{``#1''}%
\providecommand \bibnamefont  [1]{#1}%
\providecommand \bibfnamefont [1]{#1}%
\providecommand \citenamefont [1]{#1}%
\providecommand \href@noop [0]{\@secondoftwo}%
\providecommand \href [0]{\begingroup \@sanitize@url \@href}%
\providecommand \@href[1]{\@@startlink{#1}\@@href}%
\providecommand \@@href[1]{\endgroup#1\@@endlink}%
\providecommand \@sanitize@url [0]{\catcode `\\12\catcode `\$12\catcode
  `\&12\catcode `\#12\catcode `\^12\catcode `\_12\catcode `\%12\relax}%
\providecommand \@@startlink[1]{}%
\providecommand \@@endlink[0]{}%
\providecommand \url  [0]{\begingroup\@sanitize@url \@url }%
\providecommand \@url [1]{\endgroup\@href {#1}{\urlprefix }}%
\providecommand \urlprefix  [0]{URL }%
\providecommand \Eprint [0]{\href }%
\providecommand \doibase [0]{https://doi.org/}%
\providecommand \selectlanguage [0]{\@gobble}%
\providecommand \bibinfo  [0]{\@secondoftwo}%
\providecommand \bibfield  [0]{\@secondoftwo}%
\providecommand \translation [1]{[#1]}%
\providecommand \BibitemOpen [0]{}%
\providecommand \bibitemStop [0]{}%
\providecommand \bibitemNoStop [0]{.\EOS\space}%
\providecommand \EOS [0]{\spacefactor3000\relax}%
\providecommand \BibitemShut  [1]{\csname bibitem#1\endcsname}%
\let\auto@bib@innerbib\@empty
\bibitem [{\citenamefont {Schnyder}\ \emph {et~al.}(2008)\citenamefont
  {Schnyder}, \citenamefont {Ryu}, \citenamefont {Furusaki},\ and\
  \citenamefont {Ludwig}}]{schnyder2008classification}%
  \BibitemOpen
  \bibfield  {author} {\bibinfo {author} {\bibfnamefont {A.~P.}\ \bibnamefont
  {Schnyder}}, \bibinfo {author} {\bibfnamefont {S.}~\bibnamefont {Ryu}},
  \bibinfo {author} {\bibfnamefont {A.}~\bibnamefont {Furusaki}},\ and\
  \bibinfo {author} {\bibfnamefont {A.~W.}\ \bibnamefont {Ludwig}},\ }\bibfield
   {title} {\bibinfo {title} {Classification of topological insulators and
  superconductors in three spatial dimensions},\ }\href@noop {} {\bibfield
  {journal} {\bibinfo  {journal} {Phys. Rev. B}\ }\textbf {\bibinfo {volume}
  {78}},\ \bibinfo {pages} {195125} (\bibinfo {year} {2008})}\BibitemShut
  {NoStop}%
\bibitem [{\citenamefont {Kitaev}(2009)}]{kitaev2009periodic}%
  \BibitemOpen
  \bibfield  {author} {\bibinfo {author} {\bibfnamefont {A.}~\bibnamefont
  {Kitaev}},\ }\bibfield  {title} {\bibinfo {title} {Periodic table for
  topological insulators and superconductors},\ }in\ \href@noop {} {\emph
  {\bibinfo {booktitle} {AIP conference proceedings}}},\ Vol.\ \bibinfo
  {volume} {1134}\ (\bibinfo {organization} {American Institute of Physics},\
  \bibinfo {year} {2009})\ pp.\ \bibinfo {pages} {22--30}\BibitemShut {NoStop}%
\bibitem [{\citenamefont {Chiu}\ \emph {et~al.}(2016)\citenamefont {Chiu},
  \citenamefont {Teo}, \citenamefont {Schnyder},\ and\ \citenamefont
  {Ryu}}]{chiu2016classification}%
  \BibitemOpen
  \bibfield  {author} {\bibinfo {author} {\bibfnamefont {C.-K.}\ \bibnamefont
  {Chiu}}, \bibinfo {author} {\bibfnamefont {J.~C.}\ \bibnamefont {Teo}},
  \bibinfo {author} {\bibfnamefont {A.~P.}\ \bibnamefont {Schnyder}},\ and\
  \bibinfo {author} {\bibfnamefont {S.}~\bibnamefont {Ryu}},\ }\bibfield
  {title} {\bibinfo {title} {Classification of topological quantum matter with
  symmetries},\ }\href@noop {} {\bibfield  {journal} {\bibinfo  {journal}
  {Reviews of Modern Physics}\ }\textbf {\bibinfo {volume} {88}},\ \bibinfo
  {pages} {035005} (\bibinfo {year} {2016})}\BibitemShut {NoStop}%
\bibitem [{\citenamefont {Zak}(1989)}]{zak1989berry}%
  \BibitemOpen
  \bibfield  {author} {\bibinfo {author} {\bibfnamefont {J.}~\bibnamefont
  {Zak}},\ }\bibfield  {title} {\bibinfo {title} {Berry’s phase for energy
  bands in solids},\ }\href@noop {} {\bibfield  {journal} {\bibinfo  {journal}
  {Phys. Rev. Lett.}\ }\textbf {\bibinfo {volume} {62}},\ \bibinfo {pages}
  {2747} (\bibinfo {year} {1989})}\BibitemShut {NoStop}%
\bibitem [{\citenamefont {McGinley}\ and\ \citenamefont
  {Cooper}(2018)}]{mcginley2018topology}%
  \BibitemOpen
  \bibfield  {author} {\bibinfo {author} {\bibfnamefont {M.}~\bibnamefont
  {McGinley}}\ and\ \bibinfo {author} {\bibfnamefont {N.~R.}\ \bibnamefont
  {Cooper}},\ }\bibfield  {title} {\bibinfo {title} {Topology of
  one-dimensional quantum systems out of equilibrium},\ }\href@noop {}
  {\bibfield  {journal} {\bibinfo  {journal} {Phys. Rev. Lett.}\ }\textbf
  {\bibinfo {volume} {121}},\ \bibinfo {pages} {090401} (\bibinfo {year}
  {2018})}\BibitemShut {NoStop}%
\bibitem [{\citenamefont {Asb{\'o}th}\ \emph {et~al.}(2016)\citenamefont
  {Asb{\'o}th}, \citenamefont {Oroszl{\'a}ny},\ and\ \citenamefont
  {P{\'a}lyi}}]{asboth2016topo}%
  \BibitemOpen
  \bibfield  {author} {\bibinfo {author} {\bibfnamefont {J.~K.}\ \bibnamefont
  {Asb{\'o}th}}, \bibinfo {author} {\bibfnamefont {L.}~\bibnamefont
  {Oroszl{\'a}ny}},\ and\ \bibinfo {author} {\bibfnamefont {A.}~\bibnamefont
  {P{\'a}lyi}},\ }\bibfield  {title} {\bibinfo {title} {A short course on
  topological insulators},\ }\href@noop {} {\bibfield  {journal} {\bibinfo
  {journal} {Lecture notes in physics}\ }\textbf {\bibinfo {volume} {919}},\
  \bibinfo {pages} {166} (\bibinfo {year} {2016})}\BibitemShut {NoStop}%
\bibitem [{\citenamefont {Berry}(1984)}]{berry1984phase}%
  \BibitemOpen
  \bibfield  {author} {\bibinfo {author} {\bibfnamefont {M.~V.}\ \bibnamefont
  {Berry}},\ }\bibfield  {title} {\bibinfo {title} {Quantal phase factors
  accompanying adiabatic changes},\ }\href@noop {} {\bibfield  {journal}
  {\bibinfo  {journal} {Proceedings of the Royal Society of London. A.
  Mathematical and Physical Sciences}\ }\textbf {\bibinfo {volume} {392}},\
  \bibinfo {pages} {45} (\bibinfo {year} {1984})}\BibitemShut {NoStop}%
\bibitem [{\citenamefont {Alba}\ \emph {et~al.}(2011)\citenamefont {Alba},
  \citenamefont {Fernandez-Gonzalvo}, \citenamefont {Mur-Petit}, \citenamefont
  {Pachos},\ and\ \citenamefont {Garc{\'\i}a-Ripoll}}]{alba2011ToF}%
  \BibitemOpen
  \bibfield  {author} {\bibinfo {author} {\bibfnamefont {E.}~\bibnamefont
  {Alba}}, \bibinfo {author} {\bibfnamefont {X.}~\bibnamefont
  {Fernandez-Gonzalvo}}, \bibinfo {author} {\bibfnamefont {J.}~\bibnamefont
  {Mur-Petit}}, \bibinfo {author} {\bibfnamefont {J.}~\bibnamefont {Pachos}},\
  and\ \bibinfo {author} {\bibfnamefont {J.~J.}\ \bibnamefont
  {Garc{\'\i}a-Ripoll}},\ }\bibfield  {title} {\bibinfo {title} {Seeing
  topological order in time-of-flight measurements},\ }\href@noop {} {\bibfield
   {journal} {\bibinfo  {journal} {Phys. Rev. Lett}\ }\textbf {\bibinfo
  {volume} {107}},\ \bibinfo {pages} {235301} (\bibinfo {year}
  {2011})}\BibitemShut {NoStop}%
\bibitem [{\citenamefont {Atala}\ \emph {et~al.}(2013)\citenamefont {Atala},
  \citenamefont {Aidelsburger}, \citenamefont {Barreiro}, \citenamefont
  {Abanin}, \citenamefont {Kitagawa}, \citenamefont {Demler},\ and\
  \citenamefont {Bloch}}]{atala2013direct}%
  \BibitemOpen
  \bibfield  {author} {\bibinfo {author} {\bibfnamefont {M.}~\bibnamefont
  {Atala}}, \bibinfo {author} {\bibfnamefont {M.}~\bibnamefont {Aidelsburger}},
  \bibinfo {author} {\bibfnamefont {J.~T.}\ \bibnamefont {Barreiro}}, \bibinfo
  {author} {\bibfnamefont {D.}~\bibnamefont {Abanin}}, \bibinfo {author}
  {\bibfnamefont {T.}~\bibnamefont {Kitagawa}}, \bibinfo {author}
  {\bibfnamefont {E.}~\bibnamefont {Demler}},\ and\ \bibinfo {author}
  {\bibfnamefont {I.}~\bibnamefont {Bloch}},\ }\bibfield  {title} {\bibinfo
  {title} {Direct measurement of the {Zak} phase in topological bloch bands},\
  }\href@noop {} {\bibfield  {journal} {\bibinfo  {journal} {Nature Physics}\
  }\textbf {\bibinfo {volume} {9}},\ \bibinfo {pages} {795} (\bibinfo {year}
  {2013})}\BibitemShut {NoStop}%
\bibitem [{\citenamefont {Wang}\ \emph {et~al.}(2016)\citenamefont {Wang},
  \citenamefont {Xiao}, \citenamefont {Liu}, \citenamefont {Zhu},\ and\
  \citenamefont {Chan}}]{wang2016measurement}%
  \BibitemOpen
  \bibfield  {author} {\bibinfo {author} {\bibfnamefont {Q.}~\bibnamefont
  {Wang}}, \bibinfo {author} {\bibfnamefont {M.}~\bibnamefont {Xiao}}, \bibinfo
  {author} {\bibfnamefont {H.}~\bibnamefont {Liu}}, \bibinfo {author}
  {\bibfnamefont {S.}~\bibnamefont {Zhu}},\ and\ \bibinfo {author}
  {\bibfnamefont {C.~T.}\ \bibnamefont {Chan}},\ }\bibfield  {title} {\bibinfo
  {title} {Measurement of the {Zak} phase of photonic bands through the
  interface states of a metasurface/photonic crystal},\ }\href@noop {}
  {\bibfield  {journal} {\bibinfo  {journal} {Phys. Rev. B}\ }\textbf {\bibinfo
  {volume} {93}},\ \bibinfo {pages} {041415} (\bibinfo {year}
  {2016})}\BibitemShut {NoStop}%
\bibitem [{\citenamefont {Cardano}\ \emph {et~al.}(2017)\citenamefont
  {Cardano}, \citenamefont {D’Errico}, \citenamefont {Dauphin}, \citenamefont
  {Maffei}, \citenamefont {Piccirillo}, \citenamefont {de~Lisio}, \citenamefont
  {De~Filippis}, \citenamefont {Cataudella}, \citenamefont {Santamato},
  \citenamefont {Marrucci} \emph {et~al.}}]{cardano2017detection}%
  \BibitemOpen
  \bibfield  {author} {\bibinfo {author} {\bibfnamefont {F.}~\bibnamefont
  {Cardano}}, \bibinfo {author} {\bibfnamefont {A.}~\bibnamefont {D’Errico}},
  \bibinfo {author} {\bibfnamefont {A.}~\bibnamefont {Dauphin}}, \bibinfo
  {author} {\bibfnamefont {M.}~\bibnamefont {Maffei}}, \bibinfo {author}
  {\bibfnamefont {B.}~\bibnamefont {Piccirillo}}, \bibinfo {author}
  {\bibfnamefont {C.}~\bibnamefont {de~Lisio}}, \bibinfo {author}
  {\bibfnamefont {G.}~\bibnamefont {De~Filippis}}, \bibinfo {author}
  {\bibfnamefont {V.}~\bibnamefont {Cataudella}}, \bibinfo {author}
  {\bibfnamefont {E.}~\bibnamefont {Santamato}}, \bibinfo {author}
  {\bibfnamefont {L.}~\bibnamefont {Marrucci}}, \emph {et~al.},\ }\bibfield
  {title} {\bibinfo {title} {Detection of {Zak} phases and topological
  invariants in a chiral quantum walk of twisted photons},\ }\href@noop {}
  {\bibfield  {journal} {\bibinfo  {journal} {Nature communications}\ }\textbf
  {\bibinfo {volume} {8}},\ \bibinfo {pages} {1} (\bibinfo {year}
  {2017})}\BibitemShut {NoStop}%
\bibitem [{\citenamefont {Goren}\ \emph {et~al.}(2018)\citenamefont {Goren},
  \citenamefont {Plekhanov}, \citenamefont {Appas},\ and\ \citenamefont
  {Le~Hur}}]{goren2018topological}%
  \BibitemOpen
  \bibfield  {author} {\bibinfo {author} {\bibfnamefont {T.}~\bibnamefont
  {Goren}}, \bibinfo {author} {\bibfnamefont {K.}~\bibnamefont {Plekhanov}},
  \bibinfo {author} {\bibfnamefont {F.}~\bibnamefont {Appas}},\ and\ \bibinfo
  {author} {\bibfnamefont {K.}~\bibnamefont {Le~Hur}},\ }\bibfield  {title}
  {\bibinfo {title} {Topological {Zak} phase in strongly coupled lc circuits},\
  }\href@noop {} {\bibfield  {journal} {\bibinfo  {journal} {Phys. Rev. B}\
  }\textbf {\bibinfo {volume} {97}},\ \bibinfo {pages} {041106} (\bibinfo
  {year} {2018})}\BibitemShut {NoStop}%
\bibitem [{\citenamefont {Lu}\ \emph {et~al.}(2016)\citenamefont {Lu},
  \citenamefont {Schemmer}, \citenamefont {Aycock}, \citenamefont {Genkina},
  \citenamefont {Sugawa},\ and\ \citenamefont {Spielman}}]{lu2016geometrical}%
  \BibitemOpen
  \bibfield  {author} {\bibinfo {author} {\bibfnamefont {H.-I.}\ \bibnamefont
  {Lu}}, \bibinfo {author} {\bibfnamefont {M.}~\bibnamefont {Schemmer}},
  \bibinfo {author} {\bibfnamefont {L.~M.}\ \bibnamefont {Aycock}}, \bibinfo
  {author} {\bibfnamefont {D.}~\bibnamefont {Genkina}}, \bibinfo {author}
  {\bibfnamefont {S.}~\bibnamefont {Sugawa}},\ and\ \bibinfo {author}
  {\bibfnamefont {I.~B.}\ \bibnamefont {Spielman}},\ }\bibfield  {title}
  {\bibinfo {title} {Geometrical pumping with a {Bose-Einstein} condensate},\
  }\href@noop {} {\bibfield  {journal} {\bibinfo  {journal} {Phys. Rev. Lett.}\
  }\textbf {\bibinfo {volume} {116}},\ \bibinfo {pages} {200402} (\bibinfo
  {year} {2016})}\BibitemShut {NoStop}%
\bibitem [{\citenamefont {Lu}\ \emph {et~al.}(2022)\citenamefont {Lu},
  \citenamefont {Reid}, \citenamefont {Fritsch}, \citenamefont {Pi{\~n}eiro},\
  and\ \citenamefont {Spielman}}]{Lu2022Dirac}%
  \BibitemOpen
  \bibfield  {author} {\bibinfo {author} {\bibfnamefont {M.}~\bibnamefont
  {Lu}}, \bibinfo {author} {\bibfnamefont {G.~H.}\ \bibnamefont {Reid}},
  \bibinfo {author} {\bibfnamefont {A.~R.}\ \bibnamefont {Fritsch}}, \bibinfo
  {author} {\bibfnamefont {A.~M.}\ \bibnamefont {Pi{\~n}eiro}},\ and\ \bibinfo
  {author} {\bibfnamefont {I.~B.}\ \bibnamefont {Spielman}},\ }\bibfield
  {title} {\bibinfo {title} {Floquet engineering topological dirac bands},\
  }\href@noop {} {\bibfield  {journal} {\bibinfo  {journal} {arXiv preprint
  arXiv: 2202.05033}\ } (\bibinfo {year} {2022})}\BibitemShut {NoStop}%
\bibitem [{\citenamefont {Rice}\ and\ \citenamefont
  {Mele}(1982)}]{rice1982elementary}%
  \BibitemOpen
  \bibfield  {author} {\bibinfo {author} {\bibfnamefont {M.~J.}\ \bibnamefont
  {Rice}}\ and\ \bibinfo {author} {\bibfnamefont {E.~J.}\ \bibnamefont
  {Mele}},\ }\bibfield  {title} {\bibinfo {title} {Elementary excitations of a
  linearly conjugated diatomic polymer},\ }\href@noop {} {\bibfield  {journal}
  {\bibinfo  {journal} {Phys. Rev. Lett.}\ }\textbf {\bibinfo {volume} {49}},\
  \bibinfo {pages} {1455} (\bibinfo {year} {1982})}\BibitemShut {NoStop}%
\bibitem [{\citenamefont {Su}\ \emph {et~al.}(1979)\citenamefont {Su},
  \citenamefont {Schrieffer},\ and\ \citenamefont {Heeger}}]{su1979solitons}%
  \BibitemOpen
  \bibfield  {author} {\bibinfo {author} {\bibfnamefont {W.~P.}\ \bibnamefont
  {Su}}, \bibinfo {author} {\bibfnamefont {J.~R.}\ \bibnamefont {Schrieffer}},\
  and\ \bibinfo {author} {\bibfnamefont {A.~J.}\ \bibnamefont {Heeger}},\
  }\bibfield  {title} {\bibinfo {title} {Solitons in polyacetylene},\
  }\href@noop {} {\bibfield  {journal} {\bibinfo  {journal} {Phys. Rev. Lett.}\
  }\textbf {\bibinfo {volume} {42}},\ \bibinfo {pages} {1698} (\bibinfo {year}
  {1979})}\BibitemShut {NoStop}%
\bibitem [{Note1()}]{Note1}%
  \BibitemOpen
  \bibinfo {note} {Our expressions for the symmetry operations are the unitary
  part of the complete symmetry operator, where complex conjugation implicitly
  expresses the antiunitary contribution}\BibitemShut {NoStop}%
\bibitem [{\citenamefont {Cooper}\ \emph {et~al.}(2019)\citenamefont {Cooper},
  \citenamefont {Dalibard},\ and\ \citenamefont
  {Spielman}}]{cooper2019topological}%
  \BibitemOpen
  \bibfield  {author} {\bibinfo {author} {\bibfnamefont {N.~R.}\ \bibnamefont
  {Cooper}}, \bibinfo {author} {\bibfnamefont {J.}~\bibnamefont {Dalibard}},\
  and\ \bibinfo {author} {\bibfnamefont {I.~B.}\ \bibnamefont {Spielman}},\
  }\bibfield  {title} {\bibinfo {title} {Topological bands for ultracold
  atoms},\ }\href@noop {} {\bibfield  {journal} {\bibinfo  {journal} {Reviews
  of modern physics}\ }\textbf {\bibinfo {volume} {91}},\ \bibinfo {pages}
  {015005} (\bibinfo {year} {2019})}\BibitemShut {NoStop}%
\bibitem [{NuF()}]{NuFootnote}%
  \BibitemOpen
  \href@noop {} {}\bibinfo {note} {Although a winding number can be defined for
  any path that encircles $\ez$ (zero, one or many times), it is only a
  topological invariant when the path resides on the equator of the Bloch
  sphere.}\BibitemShut {Stop}%
\bibitem [{Note2()}]{Note2}%
  \BibitemOpen
  \bibinfo {note} {As discussed in the SM, when TRS and PHS are absent in the
  initial state (but CS is present) $\phi _{\protect \rm Z}$ also can evolve in
  time.}\BibitemShut {Stop}%
\bibitem [{Note3()}]{Note3}%
  \BibitemOpen
  \bibinfo {note} {Our calibrated values for $\phi _{\protect \text {rf}}$
  differed by $~3\%$ from the model prediction indicating small deviations from
  our model.}\BibitemShut {Stop}%
\bibitem [{\citenamefont {Pi{\~n}eiro}\ \emph {et~al.}(2019)\citenamefont
  {Pi{\~n}eiro}, \citenamefont {Genkina}, \citenamefont {Lu},\ and\
  \citenamefont {Spielman}}]{pineiro2019sauter}%
  \BibitemOpen
  \bibfield  {author} {\bibinfo {author} {\bibfnamefont {A.~M.}\ \bibnamefont
  {Pi{\~n}eiro}}, \bibinfo {author} {\bibfnamefont {D.}~\bibnamefont
  {Genkina}}, \bibinfo {author} {\bibfnamefont {M.}~\bibnamefont {Lu}},\ and\
  \bibinfo {author} {\bibfnamefont {I.~B.}\ \bibnamefont {Spielman}},\
  }\bibfield  {title} {\bibinfo {title} {{Sauter}--{Schwinger} effect with a
  quantum gas},\ }\href@noop {} {\bibfield  {journal} {\bibinfo  {journal} {New
  J. of Phys.}\ }\textbf {\bibinfo {volume} {21}},\ \bibinfo {pages} {083035}
  (\bibinfo {year} {2019})}\BibitemShut {NoStop}%
\bibitem [{\citenamefont {Diehl}\ \emph {et~al.}(2011)\citenamefont {Diehl},
  \citenamefont {Rico}, \citenamefont {Baranov},\ and\ \citenamefont
  {Zoller}}]{diehl2011topology}%
  \BibitemOpen
  \bibfield  {author} {\bibinfo {author} {\bibfnamefont {S.}~\bibnamefont
  {Diehl}}, \bibinfo {author} {\bibfnamefont {E.}~\bibnamefont {Rico}},
  \bibinfo {author} {\bibfnamefont {M.~A.}\ \bibnamefont {Baranov}},\ and\
  \bibinfo {author} {\bibfnamefont {P.}~\bibnamefont {Zoller}},\ }\bibfield
  {title} {\bibinfo {title} {Topology by dissipation in atomic quantum wires},\
  }\href@noop {} {\bibfield  {journal} {\bibinfo  {journal} {Nature Physics}\
  }\textbf {\bibinfo {volume} {7}},\ \bibinfo {pages} {971} (\bibinfo {year}
  {2011})}\BibitemShut {NoStop}%
\bibitem [{Note4()}]{Note4}%
  \BibitemOpen
  \bibinfo {note} {This process leaves the overall phase undefined. The choice
  of phase can be viewed as a momentum space gauge transformation which affects
  no observables, but may introduce an overall offset to the calculated Zak
  phase.}\BibitemShut {Stop}%
\bibitem [{\citenamefont {Fukui}\ \emph {et~al.}(2005)\citenamefont {Fukui},
  \citenamefont {Hatsugai},\ and\ \citenamefont {Suzuki}}]{fukui2005chern}%
  \BibitemOpen
  \bibfield  {author} {\bibinfo {author} {\bibfnamefont {T.}~\bibnamefont
  {Fukui}}, \bibinfo {author} {\bibfnamefont {Y.}~\bibnamefont {Hatsugai}},\
  and\ \bibinfo {author} {\bibfnamefont {H.}~\bibnamefont {Suzuki}},\
  }\bibfield  {title} {\bibinfo {title} {Chern numbers in discretized brillouin
  zone: efficient method of computing (spin) hall conductances},\ }\href@noop
  {} {\bibfield  {journal} {\bibinfo  {journal} {Journal of the Physical
  Society of Japan}\ }\textbf {\bibinfo {volume} {74}},\ \bibinfo {pages}
  {1674} (\bibinfo {year} {2005})}\BibitemShut {NoStop}%
\bibitem [{Note5()}]{Note5}%
  \BibitemOpen
  \bibinfo {note} {The stated uncertainties are the sample standard deviation
  of Zak phase calculated individually from about 30 separate
  measurements.}\BibitemShut {Stop}%
\bibitem [{\citenamefont {Kitagawa}\ \emph {et~al.}(2010)\citenamefont
  {Kitagawa}, \citenamefont {Berg}, \citenamefont {Rudner},\ and\ \citenamefont
  {Demler}}]{Kitagawa2010}%
  \BibitemOpen
  \bibfield  {author} {\bibinfo {author} {\bibfnamefont {T.}~\bibnamefont
  {Kitagawa}}, \bibinfo {author} {\bibfnamefont {E.}~\bibnamefont {Berg}},
  \bibinfo {author} {\bibfnamefont {M.}~\bibnamefont {Rudner}},\ and\ \bibinfo
  {author} {\bibfnamefont {E.}~\bibnamefont {Demler}},\ }\bibfield  {title}
  {\bibinfo {title} {{Topological characterization of periodically driven
  quantum systems}},\ }\href@noop {} {\bibfield  {journal} {\bibinfo  {journal}
  {Phys. Rev. B}\ }\textbf {\bibinfo {volume} {82}},\ \bibinfo {pages} {235114}
  (\bibinfo {year} {2010})}\BibitemShut {NoStop}%
\bibitem [{\citenamefont {Hsu}\ and\ \citenamefont
  {Chen}(2020)}]{hsu2020topological}%
  \BibitemOpen
  \bibfield  {author} {\bibinfo {author} {\bibfnamefont {H.-C.}\ \bibnamefont
  {Hsu}}\ and\ \bibinfo {author} {\bibfnamefont {T.-W.}\ \bibnamefont {Chen}},\
  }\bibfield  {title} {\bibinfo {title} {Topological {Anderson} insulating
  phases in the long-range {Su-Schrieffer-Heeger} model},\ }\href@noop {}
  {\bibfield  {journal} {\bibinfo  {journal} {Phys. Rev. B}\ }\textbf {\bibinfo
  {volume} {102}},\ \bibinfo {pages} {205425} (\bibinfo {year}
  {2020})}\BibitemShut {NoStop}%
\bibitem [{\citenamefont {Li}\ \emph {et~al.}(2016)\citenamefont {Li},
  \citenamefont {Duca}, \citenamefont {Reitter}, \citenamefont {Grusdt},
  \citenamefont {Demler}, \citenamefont {Endres}, \citenamefont
  {Schleier-Smith}, \citenamefont {Bloch},\ and\ \citenamefont
  {Schneider}}]{li2016bloch}%
  \BibitemOpen
  \bibfield  {author} {\bibinfo {author} {\bibfnamefont {T.}~\bibnamefont
  {Li}}, \bibinfo {author} {\bibfnamefont {L.}~\bibnamefont {Duca}}, \bibinfo
  {author} {\bibfnamefont {M.}~\bibnamefont {Reitter}}, \bibinfo {author}
  {\bibfnamefont {F.}~\bibnamefont {Grusdt}}, \bibinfo {author} {\bibfnamefont
  {E.}~\bibnamefont {Demler}}, \bibinfo {author} {\bibfnamefont
  {M.}~\bibnamefont {Endres}}, \bibinfo {author} {\bibfnamefont
  {M.}~\bibnamefont {Schleier-Smith}}, \bibinfo {author} {\bibfnamefont
  {I.}~\bibnamefont {Bloch}},\ and\ \bibinfo {author} {\bibfnamefont
  {U.}~\bibnamefont {Schneider}},\ }\bibfield  {title} {\bibinfo {title} {Bloch
  state tomography using wilson lines},\ }\href@noop {} {\bibfield  {journal}
  {\bibinfo  {journal} {Science}\ }\textbf {\bibinfo {volume} {352}},\ \bibinfo
  {pages} {1094} (\bibinfo {year} {2016})}\BibitemShut {NoStop}%
\bibitem [{\citenamefont {Fl{\"a}schner}\ \emph {et~al.}(2016)\citenamefont
  {Fl{\"a}schner}, \citenamefont {Rem}, \citenamefont {Tarnowski},
  \citenamefont {Vogel}, \citenamefont {L{\"u}hmann}, \citenamefont
  {Sengstock},\ and\ \citenamefont {Weitenberg}}]{flaschner2016Berry}%
  \BibitemOpen
  \bibfield  {author} {\bibinfo {author} {\bibfnamefont {N.}~\bibnamefont
  {Fl{\"a}schner}}, \bibinfo {author} {\bibfnamefont {B.}~\bibnamefont {Rem}},
  \bibinfo {author} {\bibfnamefont {M.}~\bibnamefont {Tarnowski}}, \bibinfo
  {author} {\bibfnamefont {D.}~\bibnamefont {Vogel}}, \bibinfo {author}
  {\bibfnamefont {D.-S.}\ \bibnamefont {L{\"u}hmann}}, \bibinfo {author}
  {\bibfnamefont {K.}~\bibnamefont {Sengstock}},\ and\ \bibinfo {author}
  {\bibfnamefont {C.}~\bibnamefont {Weitenberg}},\ }\bibfield  {title}
  {\bibinfo {title} {Experimental reconstruction of the {Berry} curvature in a
  {Floquet Bloch} band},\ }\href@noop {} {\bibfield  {journal} {\bibinfo
  {journal} {Science}\ }\textbf {\bibinfo {volume} {352}},\ \bibinfo {pages}
  {1091} (\bibinfo {year} {2016})}\BibitemShut {NoStop}%
\bibitem [{\citenamefont {Hauke}\ \emph {et~al.}(2014)\citenamefont {Hauke},
  \citenamefont {Lewenstein},\ and\ \citenamefont
  {Eckardt}}]{hauke2014tomography}%
  \BibitemOpen
  \bibfield  {author} {\bibinfo {author} {\bibfnamefont {P.}~\bibnamefont
  {Hauke}}, \bibinfo {author} {\bibfnamefont {M.}~\bibnamefont {Lewenstein}},\
  and\ \bibinfo {author} {\bibfnamefont {A.}~\bibnamefont {Eckardt}},\
  }\bibfield  {title} {\bibinfo {title} {Tomography of band insulators from
  quench dynamics},\ }\href@noop {} {\bibfield  {journal} {\bibinfo  {journal}
  {Phys. Rev. Lett.}\ }\textbf {\bibinfo {volume} {113}},\ \bibinfo {pages}
  {045303} (\bibinfo {year} {2014})}\BibitemShut {NoStop}%
\bibitem [{\citenamefont {Fl{\"a}schner}\ \emph {et~al.}(2018)\citenamefont
  {Fl{\"a}schner}, \citenamefont {Vogel}, \citenamefont {Tarnowski},
  \citenamefont {Rem}, \citenamefont {L{\"u}hmann}, \citenamefont {Heyl},
  \citenamefont {Budich}, \citenamefont {Mathey}, \citenamefont {Sengstock},\
  and\ \citenamefont {Weitenberg}}]{flaschner2018observation}%
  \BibitemOpen
  \bibfield  {author} {\bibinfo {author} {\bibfnamefont {N.}~\bibnamefont
  {Fl{\"a}schner}}, \bibinfo {author} {\bibfnamefont {D.}~\bibnamefont
  {Vogel}}, \bibinfo {author} {\bibfnamefont {M.}~\bibnamefont {Tarnowski}},
  \bibinfo {author} {\bibfnamefont {B.}~\bibnamefont {Rem}}, \bibinfo {author}
  {\bibfnamefont {D.-S.}\ \bibnamefont {L{\"u}hmann}}, \bibinfo {author}
  {\bibfnamefont {M.}~\bibnamefont {Heyl}}, \bibinfo {author} {\bibfnamefont
  {J.}~\bibnamefont {Budich}}, \bibinfo {author} {\bibfnamefont
  {L.}~\bibnamefont {Mathey}}, \bibinfo {author} {\bibfnamefont
  {K.}~\bibnamefont {Sengstock}},\ and\ \bibinfo {author} {\bibfnamefont
  {C.}~\bibnamefont {Weitenberg}},\ }\bibfield  {title} {\bibinfo {title}
  {Observation of dynamical vortices after quenches in a system with
  topology},\ }\href@noop {} {\bibfield  {journal} {\bibinfo  {journal} {Nat.
  Phys.}\ }\textbf {\bibinfo {volume} {14}},\ \bibinfo {pages} {265} (\bibinfo
  {year} {2018})}\BibitemShut {NoStop}%
\end{thebibliography}%

\newpage

\clearpage

\widetext
\begin{center}
\textbf{\large Supplemental Materials for ``Observation of dynamical topology in 1D''}
\end{center}

{\it Dynamically induced chiral symmetry breaking}
It is possible to observe a time varying Zak phase when both the initial state and the evolution Hamiltonian respect CS but not TRS. 
This scenario becomes possible by starting with eigenstates of the Rice-Mele model with complex valued tunneling matrix elements~\cite{mcginley2018topology}.
In this case CS is preserved but TRS is explicitly broken by the change between real and complex tunneling matrix elements.

\begin{figure}[t]
\includegraphics{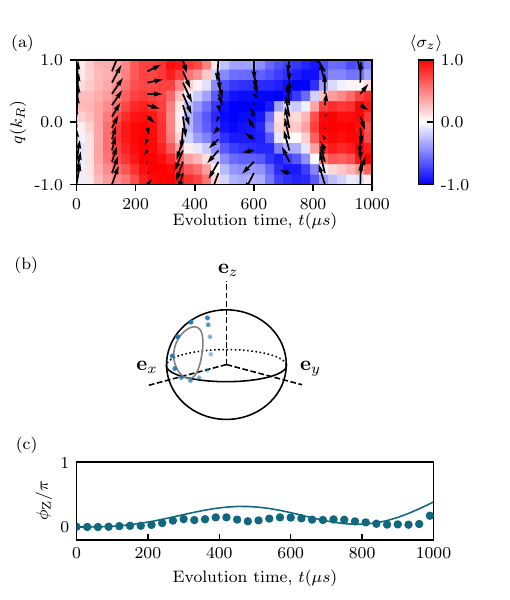}
\caption{Momentum resolved pseudospin evolution with starting $\langle \hat {\boldsymbol \sigma}(q)\rangle$ aligned along $\ey$ in a partially dimerized lattice. Model parameters were $J = 0.0362 (8) \Er$, $J^\prime = 0.1696 (5) \Er$
(a) Individual expectation values $\expx, \expy, \expz$.
As in the main text the data were filtered with root mean square Gaussian widths $\kr/8$ and $10 \us$. (b) Bloch sphere representation at $t=400 \us$
(c) Zak phase.
\label{fig_SM}} 
\end{figure}

To observe this evolution, we began with $\langle \hat {\boldsymbol \sigma}(q)\rangle=\ey$, an eigenstate of the SSH Hamiltonian with $J=0, J'=J_0 e^{\pm i \pi/2}$, and allowed it to evolve under our usual real valued Hamiltonian with $J'/J \approx 4$.
We prepared the initial state using the tunneling technique described in the main text, starting from $\ket{\downarrow}$ and evolving for $T/4$.
Figure~\ref{fig_SM}(a) shows the theoretical (left) and experimental (right) spin populations for this evolution.
The Zak phase in (b) starts at 0 and increases to $\approx \pi/6$; although the initial state breaks time-reversal symmetry, the time varying Zak phase results from the dynamically broken chiral symmetry~\cite{mcginley2018topology}.
At times beyond those shown here, the deviation between our model and the experiment increases. 
We speculate that this may correspond with the onset of dephasing from fluctuations in our control fields.
Indeed, the timescale for this process is much slower than the tunneling time because the lattice parameters used to achieve $J\approx J^\prime$ do not allow the fast tunneling rates used in the main text.

{\it Allowed changes in topological indices}
Table~\ref{app:table:topology} tabulates the different initial states that we were able to prepare, their symmetry properties and which configuration of the SSH model the evolution occurred under.
The final column describes how $\phi_{\rm Z}$ and $\nu$ can evolve in time for the given case.

\begin{table*}[b]
\begin{center}
\begin{tabular}{ |c|c|c|c|c|c|c| } 
\hline
 Initial state              & TRS   & PHS   & CS    & Symmetry class    & Configuration  & Result \\ \hline 
 Large $\Delta$ RM eigenstate  & +         & $+^*$     & $1^*$ & AI (${\rm BDI}^{*}$)            &  II (Trivial)       & $\phi_{\rm Z}$ constant \\ 
 Large $\Delta$ RM eigenstate & +         & $+^*$     & $1^*$ & AI (${\rm BDI}^{*}$)             & I (Topological)   & $\phi_{\rm Z}$ smoothly evolves \\ 
 $\phi=\pi/2$ SSH           & $+^{*}$ & $+^{*}$ & $1$   & AIII (${\rm BDI}^{*}$) & II (Trivial$^{**})$      & $\phi_{\rm Z}$ smoothly evolves \\ 
 Trivial SSH eigenstate               & $+$       & $+$       & $1$   & BDI           & I (Topological)   & $\phi_{\rm Z}$ constant, but $\nu$ jumps from $0$ to $2$ \\ 
 Topological SSH eigenstate           & $+$       & $+$       & $1$   & BDI           & II (Trivial)       & $\phi_{\rm Z}$ constant, but $\nu$ jumps from $+1$ to $-1$ \\
 \hline
\end{tabular}
\end{center}
\caption{Summary of cases studied. $^*$ denotes cases where a symmetry is present but it is different from that of the phase-free SSH model. 
$^{**}$ indicates that the evolution Hamiltonian was not fully dimerized in our experiments.}
\label{app:table:topology}
\end{table*}

\end{document}